\documentclass[aps,prl,twocolumn,nofootinbib,longbibliography]{revtex4-1}
\usepackage[babel]{csquotes}
\usepackage{graphicx}
\usepackage{amsmath,amssymb}
\usepackage[colorlinks,citecolor=blue,linkcolor=blue,urlcolor=blue]{hyperref}
\usepackage{mathrsfs}
\usepackage{enumerate}
\def\be{\begin{equation}}
\def\ee{\end{equation}}

\begin{document}

\title{Casimir cavities do not fly }

\author{Massimo Cerdonio${}^a$ and Carlo Rovelli${}^b$}
\affiliation{${}^a$ INFN Section and University, via Marzolo 8, I-35131 Padova, Italy\\
${}^b$CPT, CNRS, Aix Marseille Universit\'e 13288 Marseille and\\
Universit\'e de Toulon 7332, 83957 La Garde, France.}

\date{\small\today}

\begin{abstract}
\noindent The field inside a Casimir cavity has an effective negative mass, which acts as a buoyancy force in a gravitational field.  Can this render the total mass of the cavity negative, making it ``float" in the vacuum? Recent theoretical arguments indicate that this is impossible. We provide support to this conclusion discussing a concrete simple model of cavity, with plane parallel metallic plates kept in mechanical equilibrium by a spring and placed in a weak gravitational field.  We show that basic facts about the structure of matter imply that the total mass of the cavity is always positive.  This has implications for the hypothetical relation between vacuum energy and cosmological constant. 
\end{abstract}

\maketitle

\noindent 
It is an experimental fact that there is an attractive force between two conducting parallel plates kept at a small distance $z$ from one another. Experiments appear to fit well Casimir's theory, which predicts a force 
\be
       F_c (z) = - \frac{A}{z^4}, 
       \label{due}
\ee
where the constant $A=(\pi^2\hbar c/240)S$ depends only on the area $S$ of the plates and the fundamental constants $\hbar$ and $c$. This gives rise to a potential energy that becomes arbitrary negative as $z$ becomes small. If Casimir theory was exact, and no other phenomena came into play, the total energy of two plates at sufficiently small distance could become negative and (as simple energy conservation shows) the cavity would float upward in a gravitational field and ``fly". 

This seems implausible \cite{1,2,3}. Negative potential energy always reduces the total mass of a composed systems, but something always happens which keeps the total mass positive. For instance, the  gravitational potential energy between two gravitational masses is negative and grows as the masses approach, but before the total mass could become negative a black hole, with positive mass, forms. Recently, Bekenstein has argued on general grounds \cite{4}  that it is impossible to construct a physical system of finite size, where the Casimir energy is sufficiently negative to give a total negative mass.

So, what is wrong, or missing, with Casimir theory? The Casimir energy  stored in the cavity definitely  contributes a negative weight to the system (\cite{1,2,3,4,5,6}  and refs therein).  But Casimir theory is based on a number of idealizations which cannot be realised in nature. It is only as a consequence of these idealizations that the ``universal" aspect of \eqref{due} emerges.  Real physical systems cannot realise the ideal conditions of the Casimir model. In particular: (i) something must hold the plates at a distance, and this something stores stress and potential energy, (ii) real plates are not perfect conductors, as conductivity is a complicate physical phenomenon. Eq.\,\eqref{due} is so simple only because it depends on approximations which may not be exactly realisable in nature.  The experiment realised have checked the Casimir law in regimes where these approximations are good \cite{7}, but \eqref{due} may well be incorrect outside this regime.  

Here we explore the missing ingredients and how they affect the energy balance of a Casimir cavity. We study a more realistic model of Casmir cavity, where a spring holds the plates apart and its energy is not disregarded, and we model the plates conductivity by a plasma-Drude model \cite{8}.    We find that the crucial approximation is the infinite conductivity.  Once real properties of matter forming the plates are taken into account, the total mass of the cavity can never become negative.  Too bad, since it would have been a nice way to build aeroplanes, but real Casimir cavities do not fly. 

\vspace{.2cm}

{\em The model.} We consider two flat, parallel, conducting plates of surface area $S$ and mass $M$ each. To avoid border effects, we assume the linear dimensions of the plates to be large compared to their separation.  Successful experiments have been performed in this geometry \cite{7}; in different geometries the simple ``universal" character of \eqref{due} is lost \cite{9}. In fact it was shown long ago that cubical \cite{10} and spherical \cite{11} cavities would even give the opposite sign for the Casimir energy, not only different prefactors.We have two plates, because a single one would have vanishing Casimir energy density. This is intuitive and is in fact assumed in \cite{12}  to offer a calculation of the Casimir energy, which would intrinsically avoid the regularization of infinities.  

We assume that there is a spring between the two plates, with elastic constant $k$, which equilibrates the Casimir attraction. The system is prepared in an inertial reference frame, with the $z$ axis directed upward, the lower plate located at $z = 0$ and the upper plate at $z>0$. Let the rest position of the spring be at $z_{o} > 0$. It is convenient to normalise length unit taking $z_{o}=1$. 

The system is in equilibrium when the upper mass is at the position $z_{eq}$ where two forces are equal, that is 
\be
 -k(z_{eq}-1)-\frac{A}{z_{eq}^4}=0.
 \label{equilibrium}
\ee
This equation depends on the parameter $B=A/k>0$ which measures the ratio between Casimir and spring forces; the value $B=0$ represents a rigid cavity ($k=\infty$), while a large value of $B$ represents a soft spring.  Equation \eqref{equilibrium} has no solution for $B>B_c=4^4/5^5$ (if the spring is too soft it cannot hold the plate). It has one solution ($z_{eq}=4/5$) if $B=B_c$ and it has two solution for $B>B_c$. One of these is stable and the other is unstable.  The stable one represents the equilibrium configuration of the plate.  As one can expect, when $B$ goes to zero, namely in the limit of a very rigid spring, the stable equilibrium configuration approaches $z_0$.  

Consider the potential energy. This is the sum of two terms, due to the two forces. The Casimir attractive force \eqref{due} implies the potential energy 
\be
    E_c (z) = - \frac{A}{3z^3}
\ee
while the spring potential energy is $E_k(z) = \frac12 k (z-1)^2$.  The total potential energy is therefore
\be
    E_p (z) =  E_k(z)+E_c (z)= \frac{A}{2}  \left[\frac{(z-1)^2}B - \frac{2}{3z^3}\right].
\ee
The minimum of this energy (for the values of $B$ where it exists) is always negative; in the limit $B\!\to\! 0_-$, it approaches $E_p\!=\!-A/3$. At the other extreme, $B\to B_c$, namely a soft spring, it changes at most by about 20\%. Notice that the mass of the plates does not enter here; therefore the presence of a spring alone does not affect the possibility for the total energy to be negative. 

Let us now study how the cavity can reach equilibrium. Say it starts at large $z$, moves freely under the sole effect of the Casimir force until it touches the spring at $z=1$ and then compresses the spring to the equilibrium position $z_{eq}$.  Let $K=\frac12 M v^2$ be its kinetic energy when the plate arrives at the equilibrium point.  Then the total final energy is
\be
  E_T= \frac12 M v^2+ E_k(z)+E_{c}(z_{eq}),
\ee
and this must vanish, by conservation of energy, if the plate started at rest at large distance.
To let the system remain at the equilibrium position $z_{eq}$ , the kinetic energy $\frac12 M v^2$ must be taken away or dissipated.  Notice that to have  the  plates massless  is not truly an option, if we want to keep energy balance in this manner. Disregarding the mass entirely in this equation would mean a stable equilibrium with vanishing potential energy, which does not exist.   However, nothing prevents, so far, the mass to be arbitrarily small. 

Once the kinetic energy has been extracted, the total mass of the cavity, according to special relativity, is 
\be
   {\cal M} = [(E_k(z_{eq}) + E_{c}(z_{eq})] /c^2 + 2M 
\ee
(we disregard the mass $m_k$ of the spring).  In a gravitational field, the equilibrium position will not change \cite{13}: the forces on the upper plate are red-shifted with respect to the lower plate, but the red shift factor $(1-gz_{eq} /c^2)$ apply both to $F_k$ and $F_c $, and factorizes in the equilibrium equation. The cavity shows a net mass defect ${\cal M}-2M = E_p(z_{eq})/c^2$. There is no apparent connection between $E_p$  and the rest mass $M$ of the plates, so $M$ can be sufficiently small to make $\cal M$ negative, and the system would ``antigravitate" and ``float".  
But there is a physical issue that we have not yet considered: the conductivity of the plates. Casimir force is derived assuming the plates to be perfect conductors. Let us see what does this imply for the mass of the plates. 

We follow \cite{8}  to model conductivity. The Casimir force can also be seen as the radiation pressure exerted by vacuum fluctuations on the plates, as mirrors. The reflectivity of the metal is regulated by the plasma frequency $\omega^2_p = 4\pi e^2n/m$, where $e, m$ and $n$, are the charge, mass and volume number density of the electrons. If $\delta$ is the skin-depth for penetration of the electromagnetic field, for actual metals $\delta \sim  c/\omega_p$ for the frequencies of interest, which are those that dominate the Casimir force and are of order $c/z_{eq}$. For the plates to act as mirrors, their thickness must be larger than $\delta$, say order of $\lambda_p=2\pi c/\omega_p$.   This rules out infinitely thin plates.   There is a further correction due to the ratio between $z_{eq}$ and $\lambda_p$. Following \cite{8} , for $z_{eq}< \lambda_p$ the Casimir energy $E_{c}$ in (2) is reduced by $\sim 1.8 z_{eq} /\lambda_p$. The (negative) maximum of $E_T $ is $E_T (z_{eq})  = 1.22 E_c (z_{o})$. So for the cavity to ``float" we need $Mc^2 <\hbar E_c (z_{o})z_{eq}/\lambda_p,$ where we have inserted the reduction factor to cover the general case.  Noticing that $z_{eq} = 4/5\; z_{o}$ for maximum (negative) $E_T$, and that the plate mass is  $M = m_M n_M S \lambda_p$, with $m_M$ the atomic mass and $n_M$ the atomic volume number of the plates' metal, the ``floating" condition reads 
\be
    m_M n_M  \lambda_p^2 c^2 <  \frac{\pi^2}{720} \frac{5^2}{4^2} \frac{\hbar c}{z_{o}^2}.  
\ee
In a metal $n/n_M$ is of order one, and $m_M/m$ is $O(10^4)$, substituting for $\lambda_p^2$,  recalling that, to avoid border effect corrections, we need $z_{o} \sim (S/100)^{\frac12}$ and collecting numerical factors, we get that the surface area $S$ should satisfy the following relation between fundamental constants 
\be
   S < O(\frac{1}{720})\ \alpha  \lambda_c^2 
\ee
where $\alpha $ is the fine structure constant and  $\lambda_c$ the Compton wavelength of the electron, and we have emphasized that the surviving numerical factor is of the same order as that, with no direct physical meaning, which appears in (3). Should we release the condition $z_{eq} > \lambda_p $ and go to the opposite regime $z_{eq} < \lambda_p$, we should of course multiply the right hand side of eq (8)  by $z_{eq} /\lambda_p < 1$, making the inequality even stronger.

	Thus, equation (8) tells that it is impossible to create a cavity with parallel, flat, conducting plates which would ``antigravitate" in a weak gravitational field.
	
	One may wonder what happens in the approximation of plates of infinite electrical conductivity, an idealization best approximated by type I superconducting materials. In this case we may use the approach of ref \cite{5,6}.  Now the questionable idealization becomes the rigidity of the spring (rigidity in special and general relativity must be treated with care: an extended body in uniform acceleration, as is a Casimir cavity, develops internal stresses \cite{14,15}). We already found that the influence of the spring stiffness on $E_T (z_{eq})$ is marginal.  The maximum (negative) value for $E_T$  is easily found from the above as  $kz_{o}^2/30$. So we need $m_kc^2 < kz_{o}^2/30$, where $m_k$ is the mass of the spring, which cannot be zero. Increasing $k$ and/or $z_{o}$ may appear to be sufficient to get  ${\cal M} < 0$, but in a massive spring the compressional waves, which transmit the force between the plates separated by $z_{eq} \sim z_{o}$, have velocity $v^2 = k z_{o}^2/m_k$. Such a velocity must be lower than $c$, so $kz_{o}^2/m_k < c^2$. Then it should be $m_k < m_k/30$, which is absurd. This back of the envelope calculation is non relativistic for what concerns the vibrations of the spring, but, as for ordinary  matter the velocity of sound would be much lower than the velocity of light,  the argument  is sufficient for the present purpose to find that again Casimir cavities do not fly. 

\vspace{.2cm}

{\em  Discussion and Conclusions.} We have shown here that a realistic Casimir cavity cannot fly.  We have considered a flat parallel-plates Casimir cavity, but the result is likely to be general, for the following reason.  Eq.(8) could be weakened by a different geometry or a different nature of the material of the plates. As for the first, geometrical factors do not alter the inequality by large amounts, one or two orders of magnitude at most \cite{9}  and the inequality gets stronger when the separation between the plates gets smaller. It is therefore reasonable to expect that it holds in generic geometries. Regarding the nature of the plates, one may envision atoms much lighter than those implied in our calculation. The only possibility we see is metallic hydrogen, as the lightest metal to be possibly used in a Casimir configuration sometime in the future. In this case the term $m_M n_M$ in eq(8) would change at most again by one or two orders of magnitude. In eq (8), there is a factor of order $10^{-5}$ in front of $\lambda_c^2$, which thus appears impossible to compensate for any material in nature.  Our result is valid for generic Casimir cavities.

The Casimir effect is often presented as a ``demonstration" that the zero-point energy of a quantum field contributes to the gravity of a system, and therefore to the cosmological constant. Doubts have been raised against this idea (``�the existence of the Casimir force really doesn't say anything about the vacuum of space itself; rather, it speaks to the interactions of materials with their own nearby electromagnetic modes�" \cite{16};  ``�the Casimir effect gives no more (or less) support for the reality of the vacuum energy of fluctuating vacuum fields" \cite{17}).  The results presented here support these doubts and contradict the naive expectation for two reasons. 

First, the Casimir energy does affect the passive, and theferore, by the equivalence principle, the active gravitational mass \cite{18}.  In this sense, it "has weight". (An interesting experiment aimed at measuring directly the ``weight" of the Casimir energy, rather than the force is proposed in \cite{19}). But the mass (active and passive) of a Casimir cavity decreases when the two plates approach for the same simple reason for which the total mass of any two objects that attract one another decreases when they are approached, and their kinetic energy is taken away.  The missing energy comes from the dissipated kinetic energy, not from a mysterious vacuum source. 

Second, and more importantly, our analysis confirms that the Casimir formula is an approximation that becomes valid only after imposing unnatural idealised conditions. The attractive effect of the vacuum is inextricably connected to the interaction with the matter of the plates. This conclusion is in line with \cite{4} on general theoretical grounds, and also specifically entertained in ref \cite{20},  where �plasmonic� modes are shown to give a crucial contribution at any plate' separation. In addition, \cite{21}  points out that also actual experiments show that the sign of the Casimir energy can become positive.  Thus the structure of matter in the plates is essential, not accidental, for the force.The use of the Casimir formula in arguing for a large contribution of the free vacuum energy to the cosmological constant, is therefore unphysical.
 
Casimir cavities are not ``the demonstration" that the cosmological vacuum has weight; even less that the cosmological constant ``must" have an unreasonable value.

{\bf Acknowledements.}
	Thanks to Andrea Caranti for help in analysing the equilibrium equation, to Enrico Calloni for valuable comments and to the anonymous referees for constructive criticisms.

\end{document}